# COMPARATIVE MODEL FIDELITY EVALUATION TO SUPPORT DESIGN DECISIONS FOR COMPLEX, NOVEL, SYSTEMS OF SYSTEMS

**Edward Louis**
Clemson University
Clemson, SC

**Gregory Mocko***
Clemson University
Clemson, SC

**Evan Taylor**
Clemson University
Clemson, SC

## ABSTRACT

*Systems design processes are increasingly reliant on simulation models to inform design decisions. A pervasive issue within the systems engineering community is trusting in the models used to make decisions about complex systems. This work presents a method of evaluating the trustworthiness of a model to provide utility to a designer making a decision within a design process. Trusting the results of a model is especially important in design processes where the system is complex, novel, or displays emergent phenomena. Additionally, systems that are in the pre-prototype stages of development often do not have sources of ground truth for validating the models. Developing methods of model validation and trust that do not require real-world data is a key challenge facing systems engineers. Model fidelity in this work refers to the adherence of a model to real-world physics and is closely tied to model trust and model validity. Trust and validity directly support a designer's ability to make decisions using physics-based models. The physics that are captured in a model and the complexity of the mathematical representation of the physics contribute to a model's fidelity, and this work leverages the included physical phenomena to develop a means of selecting the most appropriate for a given design decision.*

DISTRIBUTION STATEMENT A. Approved for public release; distribution is unlimited. OPSEC9535

## 1. INTRODUCTION

Digital engineering methods are increasingly prevalent in systems engineering [1, 2]. Digital engineering refers to the process of designing systems using simulation models rather than physical prototypes. The US Army has made significant contributions to transform their acquisition and development processes, and has identified the Digital Thread as a key goal for the design and acquisition of next generation ground vehicles [1, 2, 3]. The US Army is prioritizing a reduced reliance on physical prototypes to shorten acquisition timelines and lower development and validation costs [4, 5, 6]. A challenge facing the US Army and the systems engineering community is making design decisions without a real-world source of ground truth to validate and verify the results of a simulation model against. In a systems design process where real-world data for a previously designed, analogous system does not exist, or where the development cost and timeline for a prototype of the new system is prohibitive, then design decisions must be made without results from physical testing. Instead, models are used to predict the system's performance, and real-world validation will not occur until far later in the design cycle, shown in Fig. 1 [3].

Model trust is critical in simulation-based systems design (SBSD). Systems designers must be able to trust that the results of a simulation model are leading them to the same design decisions they would make if they were designing a system with physical prototypes. In certain cases of very large, very complex systems with low production volume, no full-scale physical prototype is ever made prior to production – take aircraft carriers as an example of such a system. Means of evaluating a design decision made without ever having a source of ground truth prior to system deployment are therefore needed.

The fidelity of the models used in the simulation of not-yet-designed systems is a topic of interest for any SBSD project. Methods of evaluating model fidelity are rooted in what information a model offers to the designer facing a design decision, and why that model is preferable to any other models available to the designer [7]. Model fidelity is thus tied to the utility a model offers in a decision-making scenario: is the model at least sufficient to predict a system's performance in some testing scenario, and is it the most preferable model to use among all other sufficient models? While fidelity of a model can be assessed with respect to a model and design scenario alone, it is most pragmatic to the designer to consider a model's fidelity relative to other available models [8]. This paper discusses fidelity in the context of comparing models to predict some system performance and behavior.

*Corresponding Author
Email: gmocko@clemson.edu

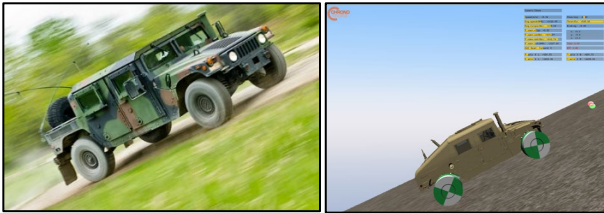

**FIGURE 1:** Real-world (left) and simulated (right) HMMWVs ascending a grade

Model fidelity refers to the extent that a model adheres to the physics and behaviors of its real-world counterpart [7, 8, 9, 10]. The mathematical representation and complexity of the included physical phenomena are key components of model fidelity. Utility metrics based on model fidelity assist the designer to select a suitable model for a design decision, ensuring that the right physical phenomena are simulated at the right level of mathematical complexity to best support selection of a design parameter. Section 2.2 presents additional background for utility in engineering decision-making contexts.

Another consideration for SBSD designers is how to proceed when the fidelity of a currently available model is not sufficient for the present design decision, and no suitable model yet exists. This prompts the designer to modify the currently available model or develop a new model of higher fidelity to address a design decision. Increasing the fidelity of a model can vary in difficulty, and two means of increasing fidelity are considered here: additional physics and behaviors are included in the model, and the current representation of the phenomena gets more mathematically complex. An example of the first type is a tire model that initially does not consider temperature in the prediction of tire-ground interactions, and is then updated to include temperature. For the second type of fidelity increase, a tire model might initially include sidewall stiffness as a constant stiffness coefficient, but later update the model to calculate sidewall stiffness as a function of the tire construction and inflation pressure.

This paper presents work done towards the identification of key components of a utility-oriented model fidelity metric, categorization of model fidelity increases, and a rigorous definition of fidelity based on the physical phenomena and features included in a model. Emphasis is placed on the difficulty of increasing the fidelity of a model. To this end, two test cases are presented in Section 4 and Section 5 to demonstrate low-effort fidelity increases and high-effort fidelity increases. A relevant example based on a military ground vehicle testing standard, TOP 02-2-610 [11] is presented in Section 6, in which three models of different fidelity are presented and represented in the proposed feature-based fidelity framework. A discussion of applying this framework to stochastic models and models developed under high levels of uncertainty is presented, as well as considerations for developing models across simulation tools in Section 7. Considerations for comparing the fidelity of models with different solver methods and convergence rates is also discussed in Section 7.

## 2. Current State of the Art
### 2.1 Simulations as a Primary Design Tool

Systems engineering is a highly complex, multi-disciplinary field concerned with the development of large-scale, interconnected, high-performing engineered systems [12, 13]. The US Army is applying cutting edge digital engineering principles in the development of its next-generation ground vehicles [2]. This is driven by the need for high performance and mobility on and off road, and cutting edge powertrain, energy storage, and autonomy technologies [14, 15]. The next generation of military ground vehicles must also have a high degree of interconnectivity with other ground vehicles and "smart" systems. This sharing of data, signals, and instructions between ground vehicles has led to an understanding of ground vehicles as systems unto themselves, as well as a systems view of multiple ground vehicles in a convoy, and ground vehicles as a system of systems across an entire theater of operations [16, 17].

The complexity and performance demands of next-generation ground vehicles and other critical systems has led to a re-evaluation of engineering design practices. Several factors drive the shift toward simulation modeling as the primary means of engineering decision-making. Among these factors is the need for fast development, acquisition, and deployment of next-generation ground vehicle technologies [2, 16, 18]. Acquisition is the process by which the US Army acquires new technologies from contractors, and the demand for shorter acquisition timelines is driving the systems engineering community to develop trustworthy simulation models to delay validation using physical prototypes in the design process [4, 6].

The systems engineering design process involves systematic reductions of the available design space, where candidate solutions are rejected from the design space if they fail some performance threshold or desired behavior. Models are used to predict how a candidate solution performs/behaves in lieu of developing and physically validating a prototype or full-production system. However, the designer must ensure both the design and the models used to inform design decisions are addressing the system's requirements [19]. In the case of digitally engineered systems, there is a significant effort to ensure that models provide valid predictions of real-world physics, since the development of a physical prototype is delayed in the design process as much as possible. Model validation is closely tied to model fidelity – the work presented here aims to tie the fidelity of a model to its validity in addressing a design decision. Given that design decisions are made on the basis of developing a valid system for a set of desired tasks, missions, and behaviors, then questions about how a model represents and considers physical phenomena become questions of a designed system's validity [8, 10].

## 2.2 Engineering Decision-Making

Design decision-making has been a topic of interest since the 1940s, when von Neumann and Morgenstern established several axioms for the behavior of rational agents faced with an adversarial, utility-maximizing "game" against similarly rational and self-interested agents [20, 21, 22]. It was shown that the behavior of a rational and self-interested agent facing a decision point involving probabilistic outcomes could be directly mapped to an optimization problem [22]. A utility-maximizing solution of the optimization problem corresponds to the analogous "best" decision in von Neumann and Morgenstern's framework. The objective function of an optimization problem is therefore a mathematical expression for a designer's preferences about system performance, and executable, physics-based models can be developed to address a designer's preferences and provide decision support in systems engineering projects.

Subsequent work in the fields of decision theory and engineering design have acknowledged that the axioms describing a rational decision-making agent are largely idealistic [23, 24]. Real human decision makers are often uncertain about factors that are critical in making an optimal decision. H. A. Simon presented a set of axioms to address the inherent inability of humans to achieve ideal rationality. Among these axioms is consideration for variability and imprecision about a human agent's preferences, personal biases, and the often qualitative nature of real-world preferences. These considerations form the basis of "bounded rationality" and these considerations prevent a neat mapping from decision theory to optimization [24]. Objective functions derived from bounded-rational preferences may therefore be "wrong" in that the optimization problem being solved is not in alignment with an ideally rational agent's preferences. Rationality also requires precise knowledge about the probability of outcomes, which may be impossible to quantify, or may be best quantified using probability density functions instead of single-valued binary probabilities. Other considerations impeding a designer from being totally rational include a lack of time to fully evaluate all choices and outcomes, and limited available computing power [24]. Model fidelity evaluation methods must account for the intractable shortcomings of human decision-makers.

Model fidelity is, as much as it is a question of validity, a question of utility provided to the designer. Model fidelity definitions and evaluation methods that cannot address the needs of a designer tasked with making a design decision are not of use to a designer of complex systems.

## 2.3 Model Fidelity Evaluation Without a Source of Ground Truth

Model fidelity is the measure of how well a model matches the behavior of real-world system [7, 10]. This is an oft-cited concept, but little consensus has been reached on the exact definition of model fidelity. Taylor et al. [8] present a set-based approach to evaluating model fidelity based on the inclusion of physical phenomena. The real world is considered to be a system of infinite detail, and the model development process involves successive reductions in set size. An instance of a simulation being executed is a finite set of inputs, assumptions, and behaviors expressed mathematically within the model containing only the phenomena relevant to a design engineer making some design decision [8].

In digital engineering contexts, design decisions are often made without a real-world source of ground truth to verify simulation results against. In some systems engineering projects, it is assumed that no ground truth will exist until the system under design enters full production [3]. Thus, any model fidelity evaluation method should provide an understanding of a model's adherence to real-world physics without an instance of the real-world system. Model fidelity evaluation is therefore not based on the accuracy of the model to the real-world, because it is not guaranteed to have a source of ground truth to compare the accuracy of the model against.

Modelers should also be wary of chasing numerical accuracy when developing surrogate models. There are several methods of determining "goodness" of a model, and the coefficient of determination is a common one for fitting a discrete set of sampled points (which may be from real life or predicted via model) and a model expressed as a continuous function [25]. Overfitting refers to choosing a mathematical expression of the surrogate on the basis of increasing its agreement with the data, without regard to the underlying physical phenomena or behavior [26]. Overfitting may yield a higher coefficient of determination between the surrogate model and the data, but the true behavior of the system is unknown, and the real-world behavior may follow a different mathematical equation than the overfitted surrogate predicts. The archetypical case is fitting a high-degree polynomial to data that is in reality logarithmic or exponential. Agreement of ground truth and model is therefore not always a trustworthy indicator of the model's fidelity – ground truth is not always available, and focusing solely on accuracy can lead to overfitting errors.

## 3. Model Fidelity Evaluation and Representation

The underlying need for a feature-based fidelity metric comes from the need to support decision-making in simulation-based systems design processes, and this support must not rely on real-world testing. Model fidelity best supports decision-making by tying the mathematical expressions and algorithms in a model to the physical phenomena and features that are most salient to a desired system behavior. For example, a vehicle's performance on grades and slopes is strongly dependent on tire mechanics, torque from the power subsystem, and location of the center of gravity [11, 27]. Grade performance is less dependent on the performance of the cooling and electrical subsystems. The validity of a model can be understood as whether a model meets the bare minimum coverage of physical phenomena to simulate some performance criterion of the vehicle, and if so, to what level of mathematical complexity and closeness to reality.

The SBSD process also includes selecting a model among possible choices in a repository. Several models may meet the requirements to simulate a parameter of interest, though some models are more difficult to set up, take more time to run, or have higher data hunger. The utility metric derived from model fidelity should also provide support in deciding from a set of suitable candidate models. In general, the most preferable model is the lowest fidelity model that is valid for a design decision while incurring the least amount of complexity, data hunger, and computational cost [7, 28]. This model utility definition is in alignment with decision-theoretic utility, where value to the decision maker is tempered by cost of executing a choice.

### 3.1 Gray Boxes
Black boxes are a way of representing behavioral models that strip a model to a set of inputs, a manipulation of those inputs that is unknown to the observer, and a set of outputs, shown in Fig. 2 [29, 30]. This representation is common in the field of engineering and computation, but has limited utility if the internal structure and behavior of a model is desired. The opposite of a black box is a white box, in which every aspect of the internal behavior of the model is visible and accessible to the user. The authors contend that a true white box is not only of limited utility, but is also an idealized model representation. If one is to accept that physics in the real-world is infinitely complex, then a truly "white" box must also be infinitely complex and contain infinite terms. All human-interpretable models must be abstracted to some finite and reduced set, therefore all boxes are to some extent gray, shown in Fig. 3. The question becomes: how gray of a box is sufficient to support the selection of a model based on the physical features represented by the model?

The authors contend that a gray box should be as "black" as possible while shedding enough light on the feature inclusion and mathematical complexity to inform a selection of model. To this end, there is a degree of subjectivity in generating a gray box. In general, one should include the least information that makes clear what inputs are taken, how they are used, and what physical phenomena are used to arrive at the variable(s) representing the system performance.

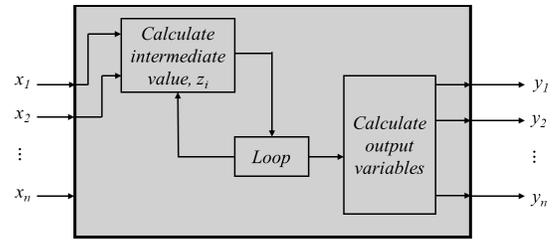

**FIGURE 3:** Gray box model showing some abstracted relationships between input and output

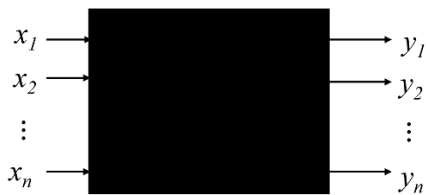

**FIGURE 2:** Black box model showing no explicit relationship between input and output

For example, a ground vehicle model that includes suspension kinematics in the prediction of some performance metric may require the 3D coordinates of all suspension hardpoints as inputs. Enumerating several dozen points just to encode the inclusion of suspension kinematics is likely too much effort for the purposes of capturing that the model includes suspension kinematics. Combining the set of suspension hardpoints into a single input of "Suspension Hardpoints" makes the gray box more interpretable for the user, and still conveys the inclusion of suspension geometric parameters. The example in Section 7 demonstrates this in practice, where several geometric parameters of a vehicle are combined into a single input in the gray box.

## 4. FIDELITY INCREASE VIA ALGEBRAIC FEATURE ADDITION
A simple case of fidelity increase is now demonstrated. In this example, a cantilevered beam with a point load, shown in Fig. 4, is modeled at two levels of fidelity. The performance metric of interest is the deflection of the beam – the designer wishes to make decisions about the beam that ensures the beam deflects no more than a desired maximum angular and vertical deflection. The lower fidelity model uses Euler-Bernoulli (EB) beam theory to predict the vertical deflection and angular deflection along the length of the beam. The model of higher fidelity includes shear as an additional phenomenon using Timoshenko-Ehrenfest (TE) beam theory. These models are presented as a very simple case of fidelity increase since the difference in mathematical formalizations is simply the superposition of extra terms on the original, low-fidelity model. The low-fidelity expression is left unchanged – the additional phenomena are included by simply adding terms onto the original mathematical formalization.

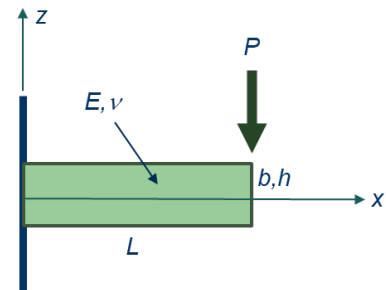

**FIGURE 4:** Cantilever beam with point load at free end

Euler-Bernoulli beam theory models the deflection of slender beams by considering only the effects of bending about the neutral axis, shown in Fig. 5. In this case, cross sections of the beam remain normal to the neutral axis. Slenderness of beams is typically evaluated using the slenderness ratio, which is given by the ratio of cross-sectional height and beam length. Beams with a slenderness ratio greater than 10 are considered slender, where the effects of bending contribute so much more to a beam's deflection that the effects of shear can be safely neglected [31].

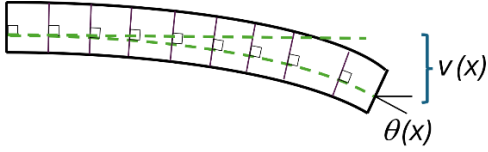

**FIGURE 5:** Euler-Bernoulli beam deflection assumptions

The mathematical formalization of the beam's curvature is a set of two polynomials that model the angular deflection and vertical deflection, respectively. The expressions are variant in $x$, the position along the beam, and considers the load, $P$, moment of inertia, $I$, Young's modulus, $E$, and beam length, $L$.

$$\theta(x) = \frac{P}{2EI}(L^2 - x^2) \tag{1}$$

$$v(x) = \frac{P}{6EI}(-x^3 + 3L^2x - 2L^3) \tag{2}$$

The higher-fidelity model, using TE beam theory, includes not only bending about the neutral axis, but also deflection due to shear loading over the face of a cross section [32]. The individual contributions of the bending terms and shear terms are superposed to predict the total deformation of the beam, as shown in Fig. 6. Because of the physical superposition of shear and bending deflection, the modeled deformation due to shear and bending on a cross section is expressed using an algebraic sum of the two terms.

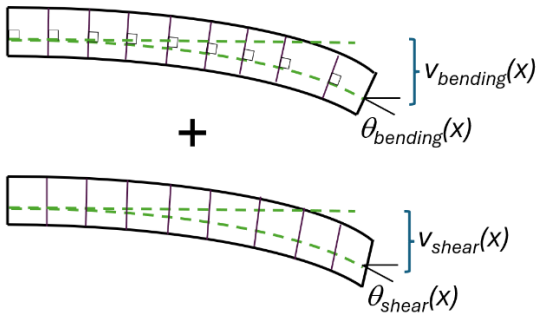

**FIGURE 6:** Timoshenko-Ehrenfest beam deflection assumptions

The TE beam model includes the original, low-fidelity EB beam model's expression for deformation due to bending, but the fidelity of a TE model is increased by including shear. This fidelity increase is a simple case because the lower fidelity expression is not modified at all; the additional shear terms are superposed onto the original EB model, which can be demonstrated by comparing Eqn. (1) and Eqn. (2) with Eqn. (3) and Eqn. (4)

$$\theta(x) = \frac{P}{\kappa AG} + \frac{P}{2EI}(L^2 - x^2) \tag{3}$$

$$v(x) = \frac{P}{6EI}(-x^3 + 3L^2x - 2L^3) - \frac{P(L-x)}{\kappa AG} \tag{4}$$

The predicted deflections of each model are shown in Figures 7 through 10. Since the TE model considers more types of deflection than the EB model, and since the TE model leaves the EB model's contributions to the beam's deflection unchanged, the TE model predicts more deflection. However, the additional deflection is most significant for short beams. The magnitude of the shear terms is far less than for bending terms at high slenderness ratios.

In the case of the simple fidelity increase, the EB model has a wide validity range – only beams below some slenderness ratio threshold, typically 8-10, are unsuitable for analysis using EB beam theory. A TE beam model predicts everything an EB model predicts for long and slender beams, but also makes valid predictions about short beams. Thus, model fidelity is not necessarily correlated to a wider range of a model's utility in supporting design decisions, though there is a tendency for cost to increase with increasing fidelity. In the beam example, the TE model requires the designer to furnish several additional pieces of information and has higher computational complexity. But the additional information and complexity of the TE model only provides utility when the beam meets several scenario-specific conditions. Otherwise, when the beam is long and slender, the validity of the TE model is the same as the EB model, but the EB model is less difficult and costly to run. Here, the additional feature inclusion of the TE formalization narrows the range of beams for which the model is useful. The additional fidelity reduces the useful range of the beam model, but it is necessary to make the change to support a design scenario where the additional physics and complexity is necessary. For short beams, the designer must deal with additional data hunger and higher computational cost.

To demonstrate this, the beam models are used to simulate a long slender beam and a short beam. Both beams have a cross-sectional height of 0.1 meters, and the long beam has a length of 1 meter, yielding a slenderness ratio of 10. The short beam has a length of 0.1 meters, yielding a slenderness ratio of 1. Fig. 7 and Fig. 8 demonstrate how the inclusion of shear in the TE model causes the prediction of deflection in a short beam to vary significantly between the models. The TE model predicts more severe deflection compared to the EB model. A designer concerned with ensuring that the designed system avoids some critical deflection value gets more decision-making utility by using the more conservative predictions from the TE model.

   

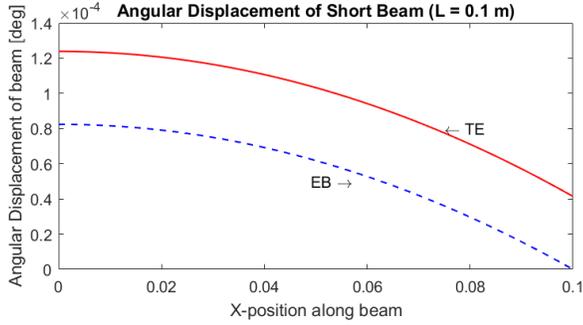

**FIGURE 7:** Angular deflection of short beam modeled with EB and TE theory

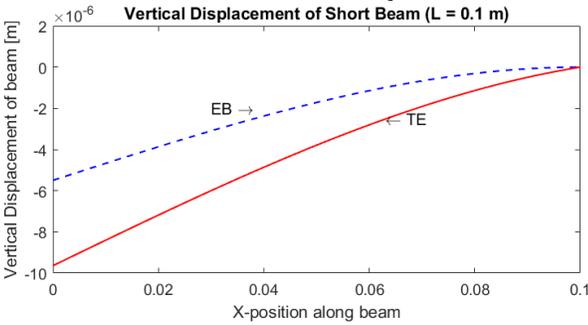

**FIGURE 8:** Vertical deflection of short beam modeled with EB and TE theory

Fig. 9 and Fig. 10 show the angular and vertical displacement for the longer beam. The curves do not vary significantly from one another. The predictive power of both models in this design setting is roughly equal, though the EB model requires fewer inputs, and incurs less computational cost.

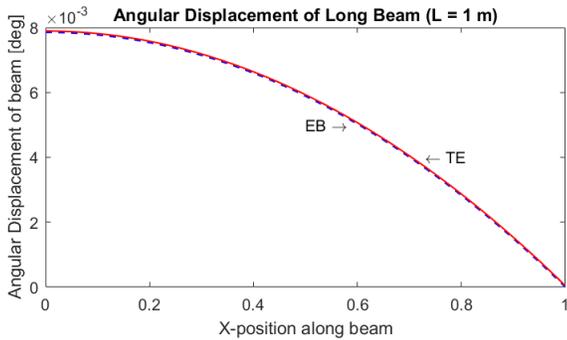

**FIGURE 9:** Angular deflection of long beam modeled with EB and TE theory

The examples show that depending on the validity frame and scenario, a lower fidelity model may be the better choice. In other cases, the higher fidelity model is a necessary choice, as the predictions made by the lower fidelity model fail to consider critical phenomena. For a long beam, the predictions of the EB and TE models are similar, and the lower cost of the EB makes it the more appropriate choice. For a short beam, the limited number of physical features in the EB model make it an invalid model, not just a less appealing model.

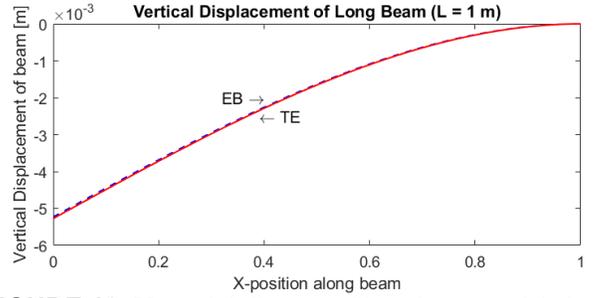

**FIGURE 10:** Vertical deflection of long beam modeled with EB and TE theory

## 5. FIDELITY INCREASE VIA MODEL REPLACEMENT

A more complex case of fidelity increase is given using the settling time of a spring-mass-damper system undergoing a step input, shown in Fig. 11. A similar approach as the beam example is taken, where a low-fidelity and high-fidelity model are compared when given the same real-world scenario. In this case, a simple heuristic model is compared to a numerical model. The settling time of a dynamic system is the time it takes for the system to reach steady state. However, a designer familiar with differential equations knows that a spring-mass-damper system settles asymptotically – it will theoretically never reach steady-state. A common criterion for determining when a system has settled is finding the time when the state variable reaches and stays within 2% of its deterministically found steady-state value.

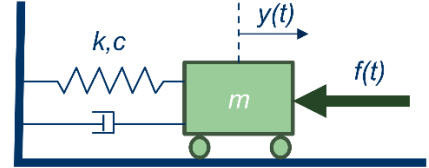

**FIGURE 11:** Spring-mass-damper schematic

A heuristic for settling time is given by Eqn. (5) which takes the spring rate, sprung mass, and damping rate and performs a simple algebraic manipulation to predict the settling time of the system by first determining the natural frequency and damping ratio. This heuristic assumes the system is strongly underdamped and is forced by a step function [33]. This assumption places a validity frame on the predictions of the model – the designer should not trust the model when evaluating overdamped systems and systems with non-step inputs.

$$t_s = \frac{4}{\zeta \omega_n} \quad (5)$$

A gray box representation of this model is given in Fig. 12, showing how the system parameters, taken as input, are manipulated into an expression for settling time.

             

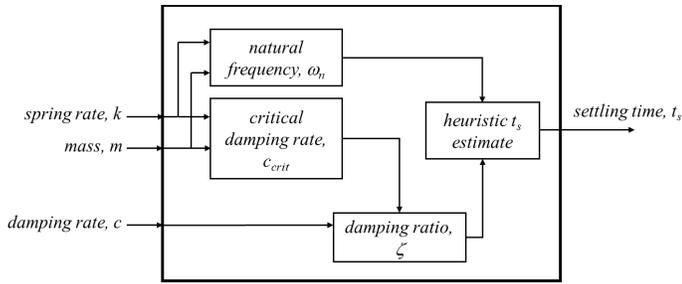
**FIGURE 12:** Gray box representation of heuristic model

The higher-fidelity model solves the ODE numerically using MATLAB's `ode23` function, taking the system parameters, initial conditions, forcing function, and settling criterion as input, shown in Fig. 13. This model places no restrictions on the damping rate of the system or the inputs it receives. However, significantly more inputs to the model are required, and the computational complexity of the model is higher. The `ode23` function numerically calculates the system's time response The model then finds the time, $t$, at which the system settles to ±2% of steady-state within the vector, $y(t)$.

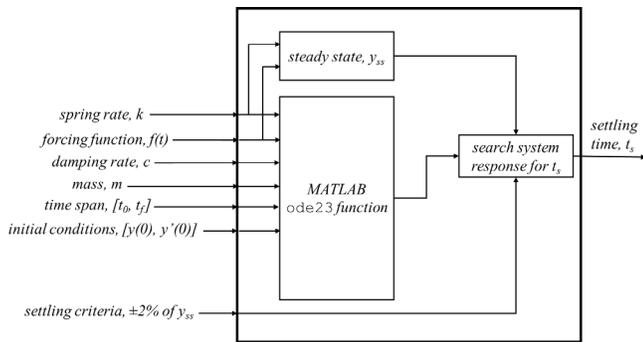
**FIGURE 13:** Gray box representation of computational model

Some of the computational model's wider validity frame can be attributed to the fact that damping ratio is calculated and used in the solution of the differential equation. The settling time of the computational model explicitly includes the damping rate's effect on energy dissipation within the system. The heuristic model calculates damping ratio, but the model never calculates how the damping ratio effects energy dissipation. The heuristic model is simply a rational function as a function of damping ratio, which diminishes towards an asymptote as damping ratio grows large. A designer's knowledge about physics and dynamic systems would tell them this is not correct, as settling time grows as the damping rate, $c$, grows large, rather than diminishing as the heuristic predicts. This exemplifies the limited validity frame of the heuristic model as a function of neglected physics and simpler mathematical manipulations within the gray box. A plot of settling time of a system over a range of damping ratios is shown in Fig. 16.

Since the system shown in Fig. 14 is underdamped, the models are in close agreement about their prediction of when the system settles. The results diverge significantly when overdamped systems are simulated, shown in Fig. 15.

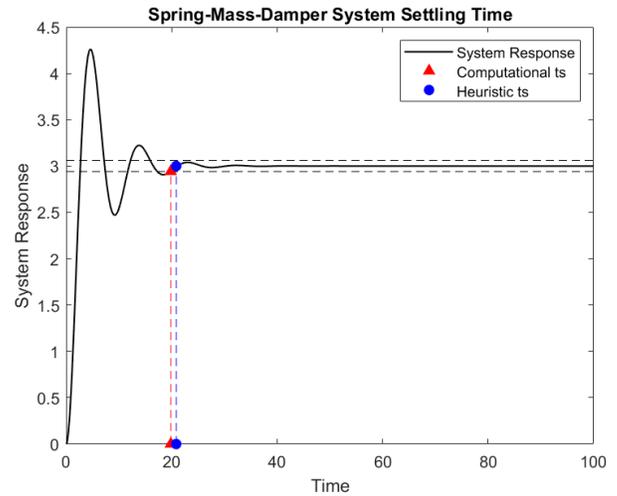
**FIGURE 14:** Settling time predictions of underdamped system

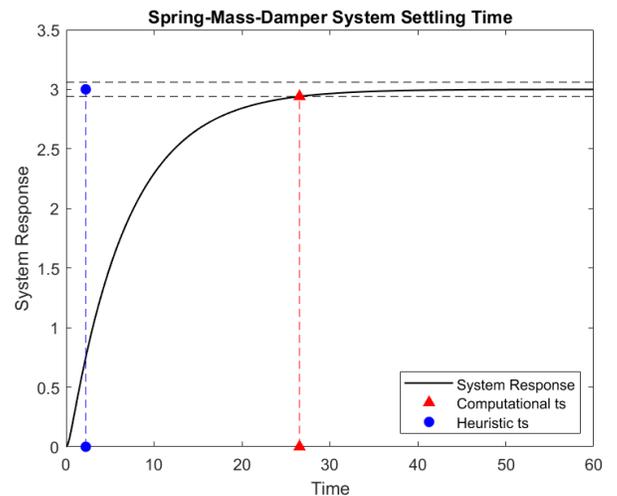
**FIGURE 15:** Settling time predictions of overdamped system

In this case, the models agree for underdamped systems, but the results diverge for overdamped systems. Increasing the level of fidelity widens the useful range of design predictions in this case. An important finding of the two test cases is that increased fidelity does not necessarily correlate with a more generally useful model. Added fidelity may only add decision-making utility to specific ranges of the validity frame, or for certain design decision points. Depending on the design decision or system behavior, increasing fidelity may make the model support a "niche" of systems or behaviors. In other cases, increasing the fidelity can make the model more applicable to a range of models.

The latter is the case for the settling time model. Fig. 16 shows the output for both models over a range of damping values. The models are overlaid in the same plot, where it is evident that the models predict similar settling times until the system is nearly critically damped. After this point, the heuristic

    

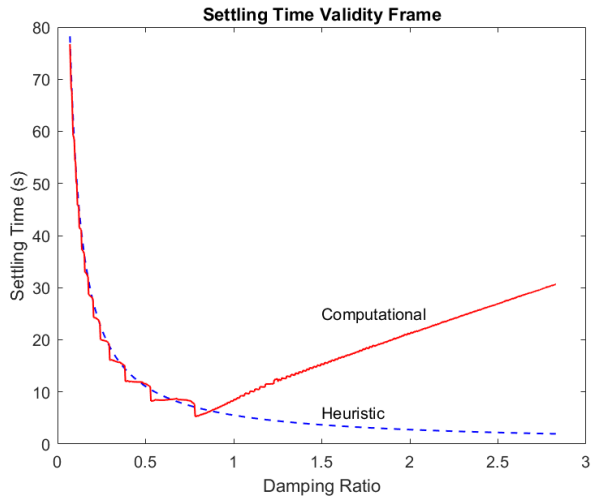

**FIGURE 16:** Comparison of settling time models over range of damping ratio

model continues to decrease toward an asymptote, while the computational model trends upwards as damping increases.

The prior two examples seek to "bound" the effort that an increase in fidelity might incur for the designer. The beam example presents a simple case, where the original model formulation is extended to include shear loads over the face of a cross section by adding polynomial terms to the original model. The settling time model presents a high-effort case, in which the low fidelity heuristic model can not be extended by adding additional terms or changing the existing terms. An entirely new model must be developed, and in this case, the new model is of considerably higher complexity, involving a numerical solution to an ODE. The next section presents a test case of three military ground vehicle models undergoing a test procedure used by the US Army to evaluate vehicle performance.

## 6. APPLICATION TO GROUND VEHICLE MODELS

Several models are developed based on a US Army Test Operations Procedure, TOP 02-2-610 Gradeability to explore differences in fidelity from the perspective of physical feature inclusion. This TOP provides instructions for testing several measures of a ground vehicle's performance on graded environments [11]. The real-world tests are conducted at the Aberdeen Proving Grounds, a US Army facility for the testing of military ground vehicles [34]. This example considers a HMMWV's performance as evaluated and measured by the procedures described in TOP 02-2-610.

This example focuses on the longitudinal gradeability test, in which a vehicle starts on a constant grade at zero initial velocity. The vehicle is evaluated on its ability to supply enough torque to the wheels to start moving up the grade, without tipping over. The output variable that indicates the vehicle's performance is the highest percent grade that can be traversed – failure to traverse is either due to insufficient tractive force, or a tip-over condition in which the front wheels lose contact with the ground. Test results of the HMMWV at Aberdeen Proving Ground were not available to the authors during the development of the models described below. The lack of real-world ground truth data to verify the models against provides an excellent test case to demonstrate model fidelity evaluation practices without relying on accuracy as a measure of fidelity. To this end, the models were developed based on testing, setup, and measurement the procedures prescribed in TOP 02-2-610, from the authors' understanding of physics, and from existing ground vehicle modeling methods, like Pacejka's Magic Formula for tire-ground interactions.

Three models at differing levels of feature inclusion were developed to simulate this testing scenario. Two were developed using MATLAB [27], and a third using Chrono [17, 35], a multi-physics simulation environment for ground vehicles. This is done to demonstrate how gray boxes and the inclusion of physical phenomena can be used to compare the fidelity of several models whether they are within the same simulation environment (comparing two MATLAB models), or across simulation tools (comparing a MATLAB model to a Chrono model).

A gray box representation for the lowest fidelity gradeability model, named RigidSusp, is shown in Fig. 17. This gray box abstracts the input space to facilitate a designer understanding the inputs, without enumerating every individual input variable and parameter. For example, rather than representing each of the four wheel center locations as four separate arrows pointing into the gray box, a single input labeled *"Location of wheel centers"* is passed to the gray box. Due to the algorithmic complexity of the model, the main functionality of the model is abstracted into a loop with two checks corresponding with a failure to ascend the grade. One loop checks for tip-over failure, the other checks if there is sufficient torque at the rear wheels to ascend the grade. The model sweeps over a range of grades until the vehicle fails in one of the conditions [27]. The last grade successfully ascended is reported.

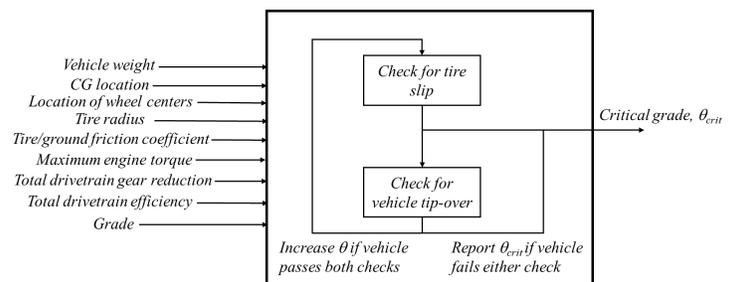

**FIGURE 17:** RigidSusp MATLAB model gray box

The second model, SpringSusp, is a modification of the first, and includes suspension kinematics as an additional phenomenon affecting performance, shown in Fig. 18. Specifically, the influence of the suspension's springs settling on the normal force on the tires is included. An additional sub-loop is added to the main loop of the algorithm. This subloop iterates until the spring displacement variables converge to their steady state position, at which point the normal force is calculated and used in the torque and tip-over conditions. Similarly to the first model, a sweep of grades is fed to the model until the tractive force condition or tip-over condition is violated [27].

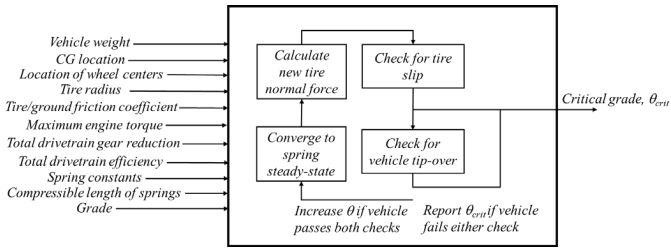

**FIGURE 18:** SpringSusp MATLAB model gray box

The third model is written in Chrono, which uses a template-based framework to model vehicle subsystems. Its gray box is shown in Fig 19. This is a time-variant system, as opposed to the quasi-static assumptions of the MATLAB models. In the Chrono model, the engine torque is determined from a lookup table based on engine RPM, and torque is applied through a transmission model that considers torque converter slip and lockup as a ramp function. Transmission shift logic is also modeled, but TOP 02-2-610 stipulates that the lowest available gear is used for the longitudinal gradeability test [11]. This is contrasted with the MATLAB models, which assume a maximum torque value applied through a global drivetrain gear reduction. The MATLAB models do not simulate the system through time, so there is no consideration for how torque is applied, a phenomenon that the Chrono model does consider.

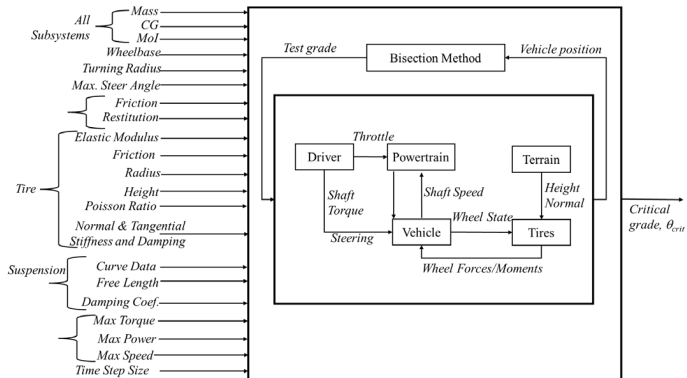

**FIGURE 19:** Chrono model gray box

The Chrono model considers rigid body suspension kinematics in its simulation of gradeability performance. A multi-link suspension subsystem model determines the position of each tire and the forces acting on the tire. This is an increase in model fidelity over the SpringSusp MATLAB model, which abstracts wheel travel to pure vertical motion that directly displaces a vertically oriented spring. In reality, the wheel travels in an arc constrained by the control arms, and the springs are mounted at some orientation not directly tangent to the path of motion of the wheel – these are kinematic considerations that the Chrono model includes and formalizes mathematically. The inputs, assumptions, and outputs of all three models are shown in Table 1.

| Model | # of Inputs | Physical Assumptions | Critical Grade |
|---|---|---|---|
| RigidSusp | 9 | Rigid suspension, rigid tire, rigid ground, Coulomb friction, no weight transfer, constant torque | 35.1% |
| SpringSusp | 11 | Suspension has constant spring rate, rigid tire, rigid ground, Coulomb friction, no weight transfer, constant torque | 35.0% |
| Chrono | 22 | Suspension has constant spring rate, meshed rigid tire, meshed rigid soil, smoothed contact model, dynamic weight transfer, dynamic torque application | 40.7% |

**TABLE 1:** Comparison of gradeability models

## 7. DISCUSSION AND FUTURE WORK

While the proposed fidelity definition is independent of simulation environment, it is noted that fidelity increases are often more difficult when the simulation environment changes. In the gradeability example, the first fidelity increase did not entail a change of simulation environment – RigidSusp and SpringSusp are both MATLAB models. While the change in modeled physics is not trivial like the beam theory example, it is not as challenging as re-evaluating the model algorithm and solution approach in a new simulation environment or modeling language.

The shift in the second gradeability model fidelity increase is a more significant step up. A gradeability simulation that considers these phenomena could be authored in MATLAB, but the author of the model noted that Chrono has many relevant subsystem models built-in because Chrono is template-based. Calling an existing high-fidelity tire model or a kinematic model is easier than writing a new one, though the overall simulation architecture must be re-evaluated and rewritten when changing simulation tool. While the simulation tool is not strictly tied to fidelity or increases in fidelity, the authors observe that fidelity increases made within the same simulation environment are often less challenging to the modeler.

The authors also contend that the convergence, numerical precision, and solution method does not affect the fidelity of the model. In the gradeability example, the MATLAB models sweep from 0% grade up to the grade at which the HMMWV can no longer ascend in some increment defined in the variable definition section of the script. The precision of the reported critical grade and computational time can be varied by orders of magnitude by changing how the critical grade value is found. The Chrono model uses the Bisection Method to determine

9  © 2025 by ASME

critical grade. The Bisection Method also requires a convergence tolerance value determined prior to execution of the simulation, but the Bisection Method is far more efficient than a fixed-step size sweep. This is not to say that the solution method of the Chrono model is why its fidelity is higher than the MATLAB model. The physics captured within the model are unchanged if the Chrono model determined critical grade using a fixed-step sweep, and likewise, the MATLAB models' fidelity would be unchanged if they converged using the Bisection Method.

All models in this paper are deterministic, though most, if not all, real-world decisions are made under uncertainty. Aleatory uncertainty does not affect the fidelity of the model. Since aleatory uncertainty applies to the intrinsic and inherent variation of inputs to a model, changing the designer's state of information about the inputs does not affect the fidelity of the model itself. A model exists at some level of fidelity, no matter the state of information the modeler has about its inputs. Presented here were several models with deterministic inputs – it is just as well to run each model in a Monte Carlo simulation. The underlying internal fidelity of the models is not affected by feeding the model deterministic inputs, probability density functions, or bounded intervals. These representations of input uncertainty do not affect the model's internal fidelity, but rather represent the state of information available to the modeler. This state of information is subject to change throughout the design process.

Quality of input information and simulation-based decision making in stochastic frameworks is a thrust of future research, as high internal model fidelity is only useful to the modeler if the state of information about inputs is sufficiently high. Consider a tire model that includes sidewall stiffness in its prediction of lateral grip – this additional feature only helps a designer if a sidewall stiffness value can be determined. A low state of information about the sidewall stiffness may take the form of a bounded interval. A high state of information about such an input might take the form of a probability density function. In this case, sufficient sampling of available data leads the designer to have a more complete understanding of the likelihood of a given stiffness value.

Epistemic uncertainty affects the fidelity of a model. Only what is known can be modeled, so phenomena that are yet undiscovered, or whose interactions with a system are yet unknown, are necessarily not included in a model. Epistemic uncertainty about real-world phenomena places an upper bound on the possible fidelity of a model, though a modeler may, and typically does, choose to neglect known phenomena when authoring a model. Designers do not often author models at the ragged edge of what is known about a system – known, modeled phenomena are often abstracted, though they could in theory be included. For example, though it is known that suspension spring stiffness affects the longitudinal gradeability performance of a ground vehicle, and models of spring settling have been developed and validated, there are still valid, useful models that neglect this phenomenon. The modeler is not experiencing an epistemic lack of knowledge about the spring rate's effect on gradeability performance, but is choosing to neglect it knowingly, out of convenience and decision-making utility. On the other hand, if it is not known how soil moisture affects gradeability, then this phenomenon cannot be modeled, and cannot be used to make decisions about gradeability.

## 8. CONCLUSION

The model fidelity definition and test cases presented here work toward a systems design framework that ties model fidelity to design decision making. Simulations are run using models that attempt to predict the real-world behavior of a not-yet designed system and allow the designer to make decisions prior to the availability of a prototype or real-world data. These predictions are necessarily imperfect, but leverage the modeler's understanding of physics and system performance. This work uses the physical phenomena that affect the performance of a vehicle undergoing some test to inform the choice of which physical phenomena to include in a model. The mathematical complexity of the representations of modeled physics is also considered – design decisions made at later stages of the design process are based on more detailed representations of the physics that contribute to the system's behavior.

In simulation-based systems design projects, the absolute fidelity of a model may be of secondary interest compared to the relative fidelity of several models. To this end, the work presented here focuses on the difference in fidelity between models predicting the same testing scenario. The difficulty of increasing fidelity is generally bounded by the examples discussed in Sections 4 and 5: a very simple case of fidelity increase is adding terms to a mathematical expression, and a very difficult case of fidelity increase is authoring a brand-new model to replace an existing low fidelity model. This evaluation of model fidelity in the context of a testing scenario – considering the physics captured and the mathematical complexity, is applied to a case of a real-world military ground vehicle's performance as predicted by three simulations of increasing fidelity.

**ACKNOWLEDGEMENTS**

DISTRIBUTION STATEMENT A. Approved for public release; distribution is unlimited. OPSEC9535

This work was supported by the Virtual Prototyping Ground Systems (VIPR-GS) Center at Clemson University and the Automotive Research Center (ARC), a US Army Center of Excellence for modeling and simulation of ground vehicles, under Cooperative Agreement W56HZV-19-2- 0001 with the US Army DEVCOM Ground Vehicle Systems Center (GVSC). All opinions, conclusions and findings wherein are those of the authors and may not be those of the affiliated institutions.